\documentclass[10pt]{article}
\usepackage{amsmath,amsfonts,amssymb}
\newcommand{\be}{\begin{equation}}
\newcommand{\ee}{\end{equation}}
\newcommand{\ba}{\begin{eqnarray}}
\newcommand{\ea}{\end{eqnarray}}

\newcommand{\la}[1]{\label{#1}}

\def\gl#1{(\ref{#1})}

\renewcommand{\thefootnote}{\fnsymbol{footnote}}

\date{}
\begin{document}
\begin{flushleft}
{\it Journal of Mathematical Sciences, Vol. 143, No.1, 2007, pp.2707-2722} \footnotetext{\tt Translated from {\it Zapiski Nauchnyikh Seminarov POMI}, Vol.335, 2006, pp.22-49 .}\\

\vskip 1cm

{\Large\bf Factorization of nonlinear supersymmetry in one-dimensional Quantum Mechanics. I: general classification of reducibility and analysis  of the third-order algebra}
\end{flushleft}
\bigskip
\setcounter{page}{2707}
\renewcommand{\thefootnote}{\arabic{footnote}}
{\bf A.A. Andrianov$^{\sharp,\diamondsuit}$ and A.V. Sokolov$^\sharp$ }\\

{$^{\sharp}$ V.A.Fock Research Institute for Physics, Sankt-Petersburg State University,198504 Russia}

{$^{\diamondsuit}$ Departament d'ECM, Universitat de Barcelona, 08028 Spain}\\
\bigskip

\begin{minipage}[1cm]{16cm}
{\it We study possible factorizations of supersymmetric (SUSY) transformations in the one-dimensional quantum mechanics into chains of elementary Darboux transformations with nonsingular coefficients. A classification of irreducible (almost) isospectral transformations and of related SUSY algebras is presented. The detailed analysis of SUSY algebras and isospectral operators is performed for the third-order case.}
\end{minipage}

\section{Introduction: definitions and notation of the SUSY QM}
\hspace*{3ex} The concept  of supersymmetric Quantum Mechanics (SUSY QM)
represents an algebraic form of transformations of (complete or partial)  spectral equivalence between different dynamical systems
\cite{coop}--\cite{fern}. At present, there is a number of reviews
\cite{genkri}--\cite{ancan} devoted to development and various applications of the SUSY QM; the reader is referred to these reviews, which are addressed for a more detailed study of this approach
to construction of isospectral systems.
Isospectral transformations of that kind are the Darboux-Moutard-Crum transformations
 \cite{darb}, \cite{crum}--\cite{matv}, which are
 are known in the theory of ordinary differential equations for a long time
\footnote{In the monograph
\cite{matv} the Darboux transformations are given for a wider class of partial differential equations including non-stationary Schr\"odinger one and some nonlinear equations.}. In the simplest cases, intertwining of two differential operators (for instance, Hamiltonians of one-dimensional quantum systems)
by means of Darboux operators entails their factorization into differential multipliers which are formed by the same Darboux operators (Schr\"odinger factorization
\cite{schr}, \cite{inf} and its generalizations \cite{abei},
\cite{suku}). However, in general, this is not the case, and both interrelation between pairs of dynamical operators ("Hamiltonians") with (almost
\footnote{We say that operators have almost equivalent spectra if their spectra are different only at a  finite number of eigenvalues.}) equivalent spectra and  structure of operators which generate the spectral equivalence are not that simple
\cite{ais} - \cite{ast0}.
Precisely this interrelation in the one-dimensional QM  is the focus of the present paper. In particular, we present rigorously justified answers to the following questions:
\quad in what cases can the higher-order Darboux-Crum transformations be constructed with the help of a sequence of intertwining transformations of lower order which relate a chain of (almost) isospectral intermediate Hamiltonians with real nonsingular
\footnote{In this case, the potentials are sufficiently smooth, but potentials having singularities weaker than
$1/x^2$ are also acceptable.} potentials; \quad
what are elementary blocks for a nonsingular factorization of intertwining
operators
; \quad  in what way is the irreducibility of elementary blocks of isospectral transformations indicated in the SUSY algebra and in the structure of kernels of those transformations? The structure of the paper is as follows. After a short reminder of notation and basic definitions of SUSY theory of isospectral transformations we formulate basic theorems on the structure of a polynomial SUSY algebra and on minimization of this algebra up to its essential part (proofs of these theorems can be found in our preceding paper
\cite{ansok}). Then we present a classification  of irreducible (almost) isospectral transformations and related SUSY algebras (partially described in \cite{ancan}, \cite{sam1}-\cite{sam3}). Next, we  define a potential class
$K$ that is invariant under transformations of the Darboux-Crum type and formulate two theorems on reducibility of differential operators of  spectral equivalence transformations. The paper is completed with a detailed analysis of the third-order SUSY algebras and isospectral operators as a first stage in proving above-mentioned theorems on reducibility. A complete proof
 will be published in a forthcoming issue.

Let us start with a definition of the SUSY algebra and notation of its components. Consider two one-dimensional Hamiltonians of the Schr\"odinger type
 $h^+=-\partial^2
+V_1(x)$ and $h^-=-\partial^2+V_2(x)$, $\partial \equiv d/dx$, which are defined on the entire axis and have nonsingular potentials
 $V_{1,2}(x)$ . We assemble the Hamiltonians into a super-Hamiltonian,
\be
H = \left(\begin{array}{cc}
h^+& 0\\
0 & h^-
\end{array}\right).  \label{hamil}
\ee
Assume that the Hamiltonians $h^+$ and $h^-$  have an (almost) equal energy spectrum of bound states and equal spectral densities of the continuous  spectrum part; let such an equivalence be provided by the Darboux-Crum \cite{darb,crum} operators $q^{\pm}_N$ with the help of intertwining,
\be
h^+ q^+_N = q^+_N h^-, \quad q^-_N h^+ = h^- q^-_N. \label{intertw}
\ee
Further on,we restrict ourselves to differential Darboux-Crum operators of finite order
 $N$,
\be
q^{\pm}_N =\sum_{k=0}^N w^{\pm}_k (x)\partial^k, \quad w^{\pm}_N
\equiv (\mp 1)^N, \label{crum}
\ee
with real, sufficiently smooth coefficients
 $w^{\pm}_k (x)$. In this case, in the fermion number representation, the nonlinear
  ${\cal N} = 1$ SUSY QM is formed by means of nilpotent supercharges,
\be
Q_N=\left(\begin{array}{cc}
 0 &  q^+_N\\
0 & 0
\end{array}\right),\quad\bar Q_N =\left(\begin{array}{cc}
 0 & 0 \\
 q^-_N & 0
\end{array}\right),\quad  Q^2_N = \bar Q^2_N = 0.
\label{char1}
\ee
Obviously, the intertwining relations
 \gl{intertw} lead to the supersymmetry of the Hamiltonian
$H$,
\be
[H, Q_N] = [H, \bar Q_N] = 0. \label{susytr}
\ee
This nonlinear SUSY algebra is closed by the following relation between the supercharges
and Hamiltonian,
\be
\{Q_N,\bar Q_N\} = P_N(H),
\label{susyN}
\ee
where  $P_N(H)$  is a differential operator of
 $2N$th order commuting with the Hamiltonian. Depending on a relation between
 the supercharges
  $Q_N,\bar Q_N$ (the intertwining operators $q^{\pm}_N$), the operator  $P_N(H)$ can be either a polynomial of the Hamiltonian if the intertwining operators are connected by the operation of transposition:
$q^{+}_N = \left(q^{-}_N\right)^t\equiv\sum_{k=0}^N(-\partial)^kw_k^-(x)$,
or a  function of both the Hamiltonian and a differential symmetry operator of odd order in derivatives (see a detailed analysis and references in
\cite{ansok}). In our paper, we confine ourselves with the first case in which the conjugated supercharge is produced by transposition, $\bar Q_N = Q^t_N $ (a relevant theorem on the structure of such a SUSY is formulated below).

\section{Basic theorems on the structure of QM with a nonlinear SUSY }

{\bf Theorem 1 (on supersymmetric algebra with transposition symmetry).}\\
{\it Let $\phi^\mp_n(x)$, $n=1$, \dots, $N$ be a basis in ${\rm{ker}}\,q_N^\mp$:
\be q_n^\mp\phi_n^\mp=0,\qquad q_N^-=(q_N^+)^t.\ee
Then:

1) the action of the Hamiltonians  $h^\pm$ on the functions $\phi_n^\pm(x)$ is described by constant $N\times N$ matrices,
\be h^\pm\phi_n^\mp=\sum\limits_{m=1}^NS_{nm}^\pm\phi_{m}^\mp\qquad
n=1,\ldots,N;\la{hs}\ee

2) the closure of the supersymmetry algebra takes a polynomial form,
\be \{Q,Q^t\}=\det[E{\bf I}-{\bf S^+}]_{E=H}=\det[E{\bf I}-{\bf S^-}]_{E=H}
\equiv P_N(H),\ee
where $\bf I$ is an identity matrix and $\bf S^\pm$ is the matrix with entries
$S_{nm}^\pm$.}

{\bf Corollary  1.} The spectra of the matrices  $\bf S^+$ and $\bf S^-$ are equal.

In what follows, for an intertwining operator, its matrix $\bf S$  is defined as the matrix which is related to operator in the same way as $\bf S^\pm$ are related to $q^\mp$. In this case, we do not specify the basis in the kernel of the intertwining operator
in which the matrix $\bf S$ is chosen if we concern only with spectral characteristics of the matrix or, that is the same, spectral characteristics of the restriction of the corresponding Hamiltonian to the kernel of the intertwining operator considered (cf.  \gl{hs}).

A basis in the kernel of the intertwining operator in which the matrix
$\bf S$ of this operator has a Jordan form  is called{\it canonical};  
elements of a canonical basis are called  {\it transformation functions}.

Assume that the intertwining operators
 $q_N^\pm$ are represented as a  product of the intertwining operators  $k_{N-M}^\pm$ and $ p_M^\pm$, $0<M<N$ so that
$$ q_N^+=p_M^+k_{N-M}^+,\quad q_N^-=k_{N-M}^- p_M^-;\qquad
p_M^+h_M=h^+ p_M^+,\quad p_M^- h^+=h_M p_M^-;$$
\be k_{N-M}^+h^-=h_M k_{N-M}^+,\quad k_{N-M}^- h_M=h^-k_{N-M}^-,\quad h_M=-\partial^2+v_M(x), \la{fakt2}\ee
where the coefficients
 $k_{N-M}^\pm$ and $p_M^\pm$
as well as the potential  $v_M(x)$ may be complex and/or singular.  The Hamiltonian $h_M$ is called  {\it intermediate } with respect to  $h^+$ and  $h^-$. In this case, by  Theorem 1, the spectrum of the matrix  $\bf S$ of the operator
$q_N^\pm$ is a union of the spectra of the matrices  $\bf S$ for the operators
$k_{N-M}^\pm$ and $p_M^\pm$.

The potentials $V_1(x)$ and $V_2(x)$ of the Hamiltonians $h^+$ and $h^-$ are interrelated by the equation
\be V_2(x)=V_1(x)-2[\ln W(x)]'',\la{v2v1}\ee
where $W(x)$ is the  Wronskian of elements of an arbitrary (a canonical as well) 
basis in ${\rm{ker}}\,q_N^-$. The validity of Eq.~\gl{v2v1} follows from the 
Liouville-Ostrogradsky relation and the equality of coefficients at  $\partial^N$ in $q_N^-h^+$ and  $h^-q_N^-$ (see the intertwining in  \gl{intertw}).

An intertwining operator $q_N^\pm$ is called {\it  minimizable } if this operator can be presented in the form
\be q_N^\pm=P(h^\pm) p_{M}^\pm=p_{M}^\pm
P(h^\mp),\la{minim}\ee
where $p_{M}^\pm$ is an operator of order  $M$ which intertwines the same Hamiltonians as  $q_N^\pm$ ( {\it i.e.} $p_{M}^\pm h^\mp=h^\pm
p_{M}^\pm$) and $P(h^\pm)$ is a polynomial of degree $(N-M)/2>0$. Otherwise the intertwining operator  $q_N^\pm$ is named as  {\it
non-minimizable}.

The following theorem contains necessary and sufficient conditions under which an intertwining operator is minimizable or not (a proof can be found in \cite{ansok}).

{\bf Theorem 2 (on minimization of an intertwining operator)} \\
{\it An intertwining operator $q_N^\pm$ can be presented in the form
\be q_N^\pm= p_{M}^\pm\prod\limits_{l=1}^m(\lambda_l-h^\mp)^{\delta k_l}, \ee
where $p_{M}^\pm$ is a nonminimizable operator intertwining the same Hamiltonians as   $q_N^\pm$ (so that $p_{M}^\pm h^\mp=h^\pm
p_{M}^\pm$),
 if and only if  a Jordan form of the matrix  $\bf S$ of the operator
$q_N^\pm$ has $m$ pairs (and no more) of Jordan cells with equal eigenvalues $\lambda_l$ such that, for the $l$-th pair,  $\delta k_l$ is  the order of the smallest cell and  $k_l+\delta k_l$ is the order  of the largest cell . In this case,
$M=N-2\sum_{l=1}^m \delta k_l = \sum_{l=1}^n k_l$, where the  $k_l$, $m+1\leqslant
l\leqslant n$ are orders of the remaining unpaired Jordan cells.}\\

{\bf Remark 1.} A Jordan form of the  matrix  $\bf S$ of the intertwining operator
$q_N^\pm$ cannot have more than two cells with the same eigenvalue
 $\lambda$; otherwise  ${\rm{ker}}(\lambda
-h^\mp)$ includes more than two linearly independent elements.

{\bf Corollary 2.} {\it Jordan forms of the matrices  $\bf S$ of the operators $q_N^+$ and $q_N^-$ coincide up to permutation of Jordan cells.}

If a Jordan form of the matrix
$\bf S$ of an intertwining operator has  cells of order higher than one, then 
the corresponding canonical bases contains not only formal solutions of the Schr\"odinger equation but also formal associated functions, which are defined as follows \cite{naim}.

A function $\psi_{n,i}(x)$ is called a {\it formal associated function of $i$-th order} of the Hamiltonian  $h$ for a spectral value
$\lambda_n$ if
\be
(h-\lambda_n)^{i+1}\psi_{n,i}\equiv0,\quad \mbox{and}\quad
(h-\lambda_n)^{i}\psi_{n,i}\not\equiv 0 . \label{canbas1}\ee
The term 'formal' emphasizes that this function is not necessarily normalizable (not necessarily belongs to $L_2(\mathbb R)$). In particular,
an associated function  $\psi_{n,0}$ of zero order is a formal eigenfunction of
 $h$ (not necessarily a normalizable solution of the homogeneous Schr\"odinger equation).

\section{Classification of really (ir)reducible SUSY trans\-for\-ma\-tions}
The intertwining operator
$q_N^\pm$ is called  ({\it really}){\it reducible} if this operator can be presented as a product of two nonsingular intertwining operators (with real coefficients)
$k_{N-M}^\pm$ and $p_M^\pm,\ 0 < M < N$ so that Eqs.
\gl{fakt2} are valid and the  intermediate Hamiltonian $h_M$ has a real nonsingular potential.
Otherwise $q_N^\pm$ is called ({\it really}){\it irreducible}.

Really irreducible, nonminimizable, intertwining operators of second order
with real coefficients can be divided into  three types
 \cite{ancan}.

{\it A really irreducible intertwining operator of I type } is a differential intertwining operator with real coefficients for which eigenvalues of the matrix
 $\bf S$ have nontrivial imaginary parts and are mutually complex conjugate.

Let us show that any intertwining operator $q_2^-$ satisfying this definition
is, in fact, really irreducible (the case of $q_2^+$ is treated similarly). Indeed,
let $\varphi^-_{1,2}(x)$ be a canonical basis of
 ${\rm{ker}}\,q_2^-$ such that
$h^+\varphi^-_{1,2}=\lambda_{1,2} \varphi^-_{1,2}$, $\lambda_1^*=\lambda_2\ne
\lambda_1$. Assume that $q_2^-$ is reducible, {\it i.e.}, there exist intertwining operators   $k_{1}^-$ and $p_1^-$ with real nonsingular coefficients such that
\be q_2^-=k_{1}^-p_1^-,\qquad p_1^-h^+=h_1p_1^-, \qquad
k_{1}^-h_1=h^-k_{1}^-,\la{fakt3}\ee
where $h_1$ is an intermediate Hamiltonian with a 
real nonsingular potential. Obviously, a basis in the kernel of  $p_1^-$  consists either of $\varphi^-_1$
or of  $\varphi^-_2$. We restrict ourselves to the case of $\varphi^-_1$ since the case of  $\varphi^-_2$ can be considered in the same manner. Then $p_1^-=\partial-(\varphi^-_1)'/\varphi^-_1$, and, consequently, $f(x)=(\varphi^-_1)'/\varphi^-_1$ is a real-valued function. But then
$$\varphi^-_1(x)=Ce^{\int f(x)\,dx}, \qquad C={\rm{Const}};$$ hence,
$$V_1(x)-\lambda_1={{(\varphi^-_1)''(x)}\over{\varphi^-_1(x)}}=f^2(x)+f'(x)$$
is a real-valued function as well, and we get a contradiction with the
condition that
 $\lambda_1^*\ne
\lambda_1$. Thus, any operator that satisfies the above definition  is indeed really irreducible.

The degenerate case $V_{2,1}(x)={\rm{Const}}$ should be singled out.
In this case, $h^+ = h^-=h_1$, and the canonical basis
 ${\rm{ker}}\,q_2^\pm$ can be chosen in the form
\be\varphi^{\pm}_1(x)=e^{kx},\qquad \varphi^\pm_2(x)=e^{k^*x},\qquad k \ne k^*\la{ee}\ee
so that eigenvalues  of the matrix $\bf S$ of the  operator $q_2^\pm$ and the operator itself are as follows:
\be
h^\mp\varphi^\pm_{1,2}=\lambda^\mp_{1,2}\varphi^\pm_{1,2},
\quad\lambda^\mp_1=V_{2,1}-k^2,\quad
\lambda^\mp_2=V_{2,1}-k^{*2},\qquad q_2^\pm=\partial^2-2{\rm{Re}}\,k\partial
+|k|^2.\la{ll}\ee

Note that potentials of the intermediate Hamiltonians which correspond to two possible factorizations of a really irreducible intertwining operator $q_2^\pm$ of the I type
 into intertwining operators of first order, {\it i.e.},
\be V_{2,1}(x)-2[\ln\varphi^\pm_{1,2}(x)]'',\la{vII}\ee
where $\varphi^+_{1,2}(x)$ ($\varphi^-_{1,2}(x)$) is a canonical basis in
${\rm{ker}}\,q_2^+$ (${\rm{ker}}\,q_2^-$), always have a nontrivial imaginary part (see
 \cite{acdi95}) with the only exception of the case 
$V_{2,1}(x)={\rm{Const}} .$

{\it A really irreducible intertwining operator of the II type } is a differential intertwining operator $q_2^\pm$  of second order with real coefficients  such that:

(1) eigenvalues  of the matrix $\bf S$ of the  operator $q_2^\pm$ are real and different;

(2) both elements  $\varphi^\pm_1(x)$ and $\varphi^\pm_2(x)$ of a canonical basis of ${\rm{ker}}\,q_2^\pm$ have zeroes.

The irreducibility of intertwining operators satisfying this definition follows from the fact that 
otherwise the equalities,
\be q_2^+=p_1^+k_{1}^+,\qquad p_1^+h_1=h^+p_1^-, \qquad
k_{1}^+h^-=h_1 k_{1}^+\la{fakt3'}\ee
take place, or, according to
\gl{fakt3}, a basis in  ${\rm{ker}}\, k_1^+$ (${\rm{ker}}\,p_1^-$) consists either of
$\varphi^\pm_1(x)$ or of $\varphi^\pm_2(x)$, and the potential of the intermediate 
Hamiltonian $h_1$ is described by one of Eqs. \gl{vII}, {\it i.e.}, has a 
singularity(ies) by the second item of the definition. We 
also note that potentials of intermediate Hamiltonians which correspond to two 
possible singular factorizations of a really irreducible intertwining operator of the 
II type into intertwining operators of first order given by \gl{vII} are 
real since the $\varphi^\pm_{1,2}(x)$can be always chosen real.

{\it A really irreducible intertwining operator of the III type } is a differential intertwining operator $q_2^\pm$ of second order with real coefficients   such that:

(1)  the eigenvalues  $\lambda_{1,2}$ of the matrix $\bf S$ of  the operator $q_2^\pm$ are equal, $\lambda_1=\lambda_2$;

(2) a canonical basis in ${\rm{ker}}\,q_2^\pm$ consists of  formal eigenfunctions,
$\varphi^\pm_{10}(x)$, and associated  functions, $\varphi^\pm_{11}(x)$, of the Hamiltonian $h^\mp$ which assemble into a Jordan cell,
$$h^\mp\varphi^\pm_{10}=\lambda_1\varphi^\pm_{10},
\qquad(h^\mp-\lambda_1)\varphi^\pm_{11}=\varphi^\pm_{10};$$

(3) $\varphi^\pm_{10}(x)$ has at least one root.

The irreducibility of an intertwining operator satisfying this definition
follows from the fact that otherwise  equalities \gl{fakt3'} take place,
or, according to
 \gl{fakt3}, a basis in ${\rm{ker}}\,k_1^+$
(${\rm{ker}}\, p_1^-$)consists of  $\varphi^\pm_{10}(x)$, and a potential of the intermediate Hamiltonian  $h_1$ is described by the equation
\be
V_{2,1}(x)-2[\ln\varphi^\pm_{10}(x)]'',\la{vIII}\ee
{\it i.e.}, has a singularity(ies) by the third item of the definition.
The potential of the intermediate Hamiltonian, which  corresponds to the only possible singular factorization of a really irreducible intertwining operator of the III type into intertwining operators of first order given by
\gl{vIII}, is real since  the $\varphi^\pm_{10}(x)$ can always be chosen real.

Obviously, other types of really irreducible nonminimizable intertwining operators of second order do not exist.\\

Further on, we formulate two assertions which characterize reducibility of intertwining operators of any order in an exhaustive way:\\

assertion (1) of Theorem 3 on the reducibility of a nonminimizable intertwining operator with real spectrum of the matrix
 $\bf S$, multiplied by an appropriate polynomial of the Hamiltonian, into (a product of) intertwining operators of first order;

assertion (2) of Theorem 4 on the reducibility of a nonminimizable intertwining operator with arbitrary spectrum of the matrix
 $\bf S$ into (a product of) intertwining operators of first order and irreducible second-order intertwining operators of the
I, II and III type.

\section{Theorems on complete reducibility of intertwining operators}

In what follows, we use a class $K$ of potentials $V(x)$ such that:

1) $V(x)$ is a real-valued function from  $C_{\mathbb R}^\infty$;

2)  there exist numbers $R_0>0$ and  $\varepsilon>0$  ($R_0$ and $\varepsilon$
depend on $V(x)$) such that the inequality  $V(x)\geqslant\varepsilon$ takes place for any $|x|\geqslant R_0$;

3) the functions
\be
\bigg(\int\limits_{\pm R_0}^x\sqrt{|V(x_1)|}dx_1\bigg)^2
\bigg({{|V'(x)|^2}\over{|V(x)|^3}}+{{|V''(x)|}\over{|V(x)|^2}}\bigg)
\ee
are bounded for  $x\geqslant R_0$ and $x\leqslant -R_0$, respectively.

In addition, we discuss  normalizability and nonnormalizability of functions at  $+\infty$ and/or at
 $-\infty$; these properties are defined as follows.

A function $f(x)$ is called {\it normalizable at $+\infty$ (at $-\infty$)} if there exists
a real number $a_+$ ($a_-$) such that
\be
\int\limits_{a_+}^{+\infty}|f(x)|^2\,dx<+\infty\qquad
\bigg(\int\limits_{-\infty}^{a_-}|f(x)|^2\,dx<+\infty\bigg).\ee
Otherwise $f(x)$ is called {\it nonnormalizable at $+\infty$
(at $-\infty$).}

{\bf Theorem 3. (on reducibility of "dressed" nonminimizable intertwining operators)} \\
{\it Assume that the following conditions are satisfied: 

1) $h^+=-\partial^2+V_1(x)$, $V_1(x)\in K$,
and the potential $V_2(x)$ of the Hamiltonian $h^-$ is real and continuous;

2) $h^+$ and $h^-$ are intertwined by a nonminimizable differential operator of
$N$th order $q_N^-$ with coefficients from $C^2_{\mathbb R}$, so that
\be q_N^-h^+=h^-q_N^-;\ee

3) the algebraic multiplicity of $\lambda_i$, the  $i$th eigenvalue of the matrix $\bf
S$ for the operator $q_N^-$,  is equal to $k_i$, $i=1$, \dots, $n$, so that $k_1+\dots+k_n
=N$; all of the numbers  $\lambda_i$ are real and satisfy the inequalities
\be 0\geqslant\lambda_1>\lambda_2>\ldots>\lambda_n;\ee

4) $\Lambda$ is the spectrum of the matrix  $\bf S$ of the operator $q_N^-$;

5) $E_{i\pm}$, $i=0$, $1$, $2$, \dots , is the  energy of  the $i$th (from below) bound state of
$h^\pm$; $N^\pm$ is the  number of bound states of $h^\pm$ with energies of which are included into $\Lambda$; $N_\pm$ be a number of bound states of  $h^\pm$ with energies not exceeding
 $\lambda_1$;

6) \be P_\pm(E)=\prod\limits_{E_{i\pm}<\lambda_1,E_{i\pm}\not\in\Lambda}
(E-E_{i\pm}).\ee

Then: 1) $V_2(x)\in K$; coefficients of $q_N^-$ belong to $C^\infty_{\mathbb
R}$ and are real; $q_N^+=(q_N^-)^t$ has real coefficients from $C^\infty_{\mathbb R}$ and intertwines $h^+$ and $h^-$, so that
\be h^+q_N^+=q_N^+h^-;\la{splet}\ee

2) $P_+(E)\equiv P_-(E)$; the degree of $P_\pm(E)$ is equal to $N_+-N^+=N_--N^-$;

3) the operator $q_N^\mp P_\pm(h^\pm)$ intertwines $h^+$ and $h^-$ and can be
presented as a product of $N+N_++N_--N^+-N^-$ intertwining operators of first order with real coefficients from
$C^\infty_{\mathbb R}$, so that:

a) potentials of all the intermediate Hamiltonians belong to  $K$;

b) the  eigenvalue of the matrix  $\bf S$ of the $l$-th  operator (from the right) in the
factorization under consideration is equal to  $E_{l-1,\pm}$, $l=1$, \dots, $N_\pm$ and an element of the kernel of this operator is normalizable at both infinities;

c) the eigenvalue of the matrix  $\bf S$ of the $l$-th  operator (from the left) in the
factorization under consideration is equal to  $E_{l-1,\mp}$, $l=1$, \dots, $N_\mp$ and an element of the kernel of this operator is nonnormalizable at both infinities;

d) the set of eigenvalues of the matrices  $\bf S$ for operators from the $N_\pm+1$-th to the
$N_\pm+N-N^+-N^-$-th  one (from the right) in the factorization under consideration
coincides with\footnote{In this formula, one has to take into account multiplicities of eigenvalues
as follows: if  $\lambda$ is contained in $\Lambda$ with algebraic multiplicity  $K_1$, in
$\{E_{i+}\}$ with multiplicity $K_2$ and in $\{E_{i-}\}$ with multiplicity $K_3$ (obviously, $K_2$
and $K_3$  can take values  0 and 1 only), then the value $\lambda$ is contained in
$\Lambda\setminus(\{E_{i+}\}\cup\{E_{i-}\})$ with multiplicity
 $K_1-K_2-K_3$ if $K_1>K_2+K_3$ or is not contained if
$K_1\leqslant K_2+K_3$.} $\Lambda\setminus(\{E_{i+}\}\cup\{E_{i-}\})$. In addition, the eigenvalue of the matrix
$\bf S$ for an operator of this group does not decrease as the number of the operator increases (from the right to  left); a basis element of the kernel of any operator in this group is normalizable at one of the infinities only.}

{\bf Theorem 4. (on complete reducibility of nonminimizable intertwining operators)} \\ {\it Assume that the following conditions are satisfied:

 1) $h^+=-\partial^2+V_1(x)$, $V_1(x)\in K$;
the potential $V_2(x)$ of the Hamiltonian $h^-$ is real and continuous;

2) $h^+$ and $h^-$ are intertwined by a nonminimizable differential operator $q_N^-$ of
$N$th order with real coefficients from  $C^2_{\mathbb R}$, so that

\be q_N^-h^+=h^-q_N^-;\ee

3) the algebraic multiplicity of $\lambda_i$, the $i$th eigenvalue of matrix $\bf
S$ for operator $q_N^-$, is equal $k_i$, $i=1$, \dots, $n$, so that $k_1+\dots+k_n
=N$; the set of values $\lambda_i$ contains $M$ real values and $L$ pairs of mutually complex conjugate ones, so that $M+2L=n$;
the numbers $i=1$, \dots, $M$ correspond to real $\lambda_i$, and $\lambda_i>\lambda_{i+1}$,
$i=1$, \dots, $M-1$;

4) if $\lambda_1$ is real, then $\lambda_1\leqslant0$;

5) $E_{i\pm}$, $i=0$, $1$, $2$, \dots is the energy of the $i$th  bound state (from below) of
$h^\pm$; $K_\pm=\max\{i:\lambda_i>E_{0\pm}\}$, if $\lambda_1>E_{0\pm}$, and
$K_\pm=0$, if either $\lambda_1\leqslant E_{0\pm}$ or ${\rm{Im}}\,\lambda_1\ne0$.

Then: 1) $V_2(x)\in K$; coefficients of $q_N^-$ belong to $C^\infty_{\mathbb
R}$; $q_N^+=(q_N^-)^t$ has real coefficients from  $C^\infty_{\mathbb
R}$ and intertwines $h^+$ and $h^-$, so that
\be h^+q_N^+=q_N^+h^-;\ee

2) $q_N^{\mp}$ can be presented as a product of
really irreducible intertwining operators of first and second order with real coefficients from  $C^\infty_{\mathbb R}$, so that:

a) potentials of all the intermediate Hamiltonians belong to  $K$;

b) the first
\be J_1=\sum\limits_{i=M+1}^{M+L}k_i\ee
operators from the right in the factorization of $q_N^\mp$ under consideration have an order $2$ and are really irreducible operators of the I type
; in addition, one can realize that the related pairs of mutually complex conjugated eigenvalues of the matrix $\bf S$ operator $q_N^-$ are ordered arbitrarily;

c) the second (from the right) group of operators in the factorization under consideration consists of
\be J_{2\mp}=N-2J_1-2J_{3\mp},\ee
operators of first order,
where
\be J_{3\mp}=\Big[{1\over2}\sum\limits_{i=1}^{K_\mp}k_i\Big],\ee
and

(i) if  $\sum\limits_{i=1}^{K_\mp}k_i$ is even, then the eigenvalue of the matrix  $\bf S$ for the operator $q_N^-$ which corresponds to the  $l$th (from the right) of these operators does not exceed the eigenvalue related to the  $l+1$-th operator, $l=1$, \dots, $J_{2\mp}-1$;

(ii) if   $\sum\limits_{i=1}^{K_\mp}k_i$ is odd, then the eigenvalue of the matrix $\bf S$ for the operator $q_N^-$ which corresponds to  the $l$th (from the right) of these operators  does not exceed the eigenvalue related to the  $(l+1)$th operator, $l=1$, \dots, $J_{2\mp}-2$; $\lambda_{K_\mp}$ is an eigenvalue of the $(J_{2\mp}-1)$th operator and $\lambda_{K_\mp+1}$ is an eigenvalue of the  $(J_{2\mp})$th operator; in this case, the latter eigenvalue is equal
to $E_{0\mp}$;

d) the third (from the right) and the last group of operators in the factorization under consideration consists of   $J_{3\mp}$ really irreducible operators of II and III type, wherein the largest of eigenvalues of the matrix  $\bf S$ for the operator  $q_N^-$ which corresponds to the
 $l$th of these operators (from the right) does not exceed the smallest eigenvalue of the matrix  $\bf S$ for the operator $q_N^-$ which corresponds to the
 $(l+1)$th of these operators, $l=1$, \dots, $J_{3\mp}-1$.}

{\bf Remark 1.} If $E_{0\mp}$ is not an eigenvalue of the matrix $\bf
S$ for the operator $q_N^-$, then $\sum\limits_{i=1}^{K_\mp}k_i$ is even since otherwise the eigenvalue of  $q_N^\mp
q_N^\pm\equiv\prod_{i=1}^n(h^\mp-\lambda_i)^{k_i}$ at the ground state wave function of $h^\mp$ is negative.

Proofs of these theorems will be published in a forthcoming issue.\\

The next aim of this paper  is to show rigorously that any nonminimizable intertwining operator
{\it of third order} with real coefficients is really reducible \footnote{For preliminary consideration  of this issue, see in \cite{acin1,iofnis}}. Such a proof is a necessary stage in the study of reducibility of
intertwining operators of arbitrary order. For this purpose, we first derive differential equations for Wronskians  of subsets of a canonical basis of the kernel of the intertwining operator $q_N^-$ of an arbitrary order $N$. These equations form a base of the proof of reducibility for an arbitrary
intertwining operator of third order (Theorem 5) and also can be used, for instance, to examine reducibility of intertwining operator in the general case where  a Jordan form of its matrix
 $\bf S$ is a single Jordan cell.

\section{Derivation of system of equations for partial Wronskians}

Let $\phi_j(x)$, $j=1$, \dots, $N$ be a canonical basis in  ${\rm{ker}}\,
q_N^-$ and let $\lambda_j$, $j=1$, \dots, $N$ be an eigenvalue of the matrix $\bf S$
for the  operator $q_N^-$ corresponding to the Jordan cell to which  $\phi_j(x)$  is related. It is shown in \cite{ansok}, Lemma 1,   that:

1) the intertwining operator  $q_N^-$ can be presented as follows
\be q_N^-=r_1^-\ldots r_N^-, \la{l11}\ee
where the Darboux operators
\be r_j^-=\partial+\chi_j(x),\qquad j=1,\ldots,N,\la{l12}\ee
can be chosen to satisfy the equalities
\begin{equation}r_j^-\ldots r_N^-\phi_{j}=0,\qquad j=1,\ldots,N;
\la{alpha}\end{equation}

2) the following relations take place
$$ (r_{j}^-)^tr_{j}^-+\lambda_{j}=r_{j+1}^-(r_{j+1}^-)^t+\lambda_{j+1}\equiv
h_j,\qquad j=1,\ldots,N-1,$$
\be (r_N^-)^tr_N^-+\lambda_N=h^+\equiv h_N,\qquad
r_1^-(r_1^-)^t+\lambda_1=h^-\equiv h_0;\la{aaaprom}\ee

3) the intermediate Hamiltonians  $h_j$, $j=1$, \dots, $N-1$ have the Schr\"odinger
 form:
\be h_j=-\partial^2+v_j(x),\qquad  v_j(x)=\chi^2_{j}(x)-\chi'_{j}(x)+
\lambda_{j}=\chi^2_{j+1}(x)+\chi'_{j+1}(x)+\lambda_{j+1},\ee
\begin{equation} V_1(x)\equiv
v_N(x)=\chi^2_{N}(x)-\chi'_{N}(x)+\lambda_{N},\qquad V_2(x)\equiv v_0(x)=
\chi^2_{1}(x)+\chi'_{1}(x)+\lambda_{1},
\la{beta}\end{equation}
but in general with complex and/or singular potentials;

4) the intertwining relations 
\be h_lr_{l+1}^-=r_{l+1}^-h_{l+1},\qquad (r_{l+1}^-)^th_l=h_{l+1}
(r_{l+1}^-)^t,\qquad l=0,\ldots,N-1.\la{aaaspl}\ee
are realized.

Let us introduce generalized Crum determinants
\be W_j(x)=\begin{vmatrix}\phi_N(x)&\phi'_N(x)&\ldots&\phi_N^{(N-j)}(x)\\
\phi_{N-1}(x)&\phi'_{N-1}(x)&\ldots&\phi_{N-1}^{(N-j)}(x)\\ \hdotsfor{4}\\
\phi_{j}(x)&\phi'_{j}(x)&\ldots&\phi_{j}^{(N-j)}(x)\\
\end{vmatrix},\qquad j=1,\ldots,N. \ee
By \gl{alpha}, the following expressions for intertwining operators $r_j^-\ldots r_N^-$ are valid:
\be r_j^-\ldots r_N^-={1\over W_{j}(x)}
\begin{vmatrix}\phi_N(x)&\phi'_N(x)&\ldots&\phi_N^{(N-j+1)}(x)\\
\hdotsfor{4}\\
\phi_{j}(x)&\phi'_{j}(x)&\ldots&\phi_{j}^{(N-j+1)}(x)\\
1&\partial&\ldots&\partial^{N-j+1}\end{vmatrix},\qquad j=1,\ldots,N\la{rjrn}\ee
therefore,
\be r_j^-\ldots r_N^-\phi_{j-1}={{W_{j-1}(x)}\over{W_j(x)}},\qquad j=2,
\ldots,N. \la{aaaos}\ee
Furthermore, the intermediate superpotentials $\chi_j(x)$ are as follows:
\be \chi_j(x)=-{{[W_j(x)/W_{j+1}(x)]'}\over{W_j(x)/W_{j+1}(x)}}=
-{{W'_j(x)}\over{W_j(x)}}+{{W'_{j+1}(x)}\over{W_{j+1}(x)}},\qquad j=1,
\ldots,N,\quad W_{N+1}(x)\equiv1.\la{aaasuper}\ee

To obtain differential equations satisfied by the Wronskians $W_j(x)$,
we convert the expression
\be q_j^-(q_j^-)^t{{W_{j-1}(x)}\over{W_j(x)}},\qquad j=2,\ldots,N,\la{aaa1}\ee
in two different ways. On the one hand, we take into account
 \gl{aaaprom}, \gl{aaaos} and intertwinings \gl{aaaspl} to show that
\be q_j^-(q_j^-)^t{{W_{j-1}(x)}\over{W_j(x)}}\!=\!(h_{j-1}-\lambda_j)
q_j^-\ldots q_N^-\phi_{j-1}\!=\!q_j^-\ldots q_N^-(h^+-\lambda_j)\phi_{j-1}
\!=\!(\lambda_{j-1}-\lambda_{j}){{W_{j-1}(x)}\over{W_j(x)}}, \label{44} \ee
Obviously, equalities \eqref{44} are valid not only if
$\phi_{j-1}$ is a formal eigenfunction of   $h^+$ but also if
$\phi_{j-1}$ is a formal associated function of $h^+$. On the other hand, Eq.
\gl{aaa1} can be transformed as follows,
$$q_j^-\Big[-\partial-{{W'_{j}}\over{W_{j}}}+{{W'_{j+1}}\over{W_{j+1}}}\Big]
{{W_{j-1}(x)}\over{W_j(x)}}\!=
\!q_j^-\Big[-{{W'_{j-1}}\over{W_{j}}}+{{W_{j-1}W'_j}\over{W^2_j}}-
{{W'_jW_{j-1}}\over{W^2_{j}}}+{{W'_{j+1}W_{j-1}}\over{W_{j+1}W_j}}\Big]$$
$$=-\Big[\partial-{{W'_{j}}\over{W_{j}}}+{{W'_{j+1}}\over{W_{j+1}}}\Big]
\Big[{{W_{j+1}}\over{W_{j}}}\Big({{W_{j-1}}\over{W_{j+1}}}\Big)'\Big]=
-2\Big({{W_{j+1}}\over{W_j}}\Big)'\Big({{W_{j-1}}\over{W_{j+1}}}\Big)'-
{{W_{j+1}}\over{W_j}}\Big({{W_{j-1}}\over{W_{j+1}}}\Big)''$$
\be=-{{W_{j}}\over{W_{j+1}}}
\Big[\Big({{W_{j+1}}\over{W_j}}\Big)^2
\Big({{W_{j-1}}\over{W_{j+1}}}\Big)'\,\Big]',\ee
where we use \gl{aaasuper}.
Finally we obtain the equations,
\be\Big[\Big({{W_{j+1}}\over{W_j}}\Big)^2
\Big({{W_{j-1}}\over{W_{j+1}}}\Big)'\,\Big]'+(\lambda_{j-1}-\lambda_j)
\Big({{W_{j+1}}\over{W_j}}\Big)^2{{W_{j-1}}\over{W_{j+1}}}=0,
\qquad j=2, ..., N .\la{aaa2}\ee
For  further purposes,  it is convenient to introduce the functions
$w_j=W'_j/W_j$,
\be w'_j-w'_{j+2}+w^2_j-w^2_{j+2}-2w_{j+1}(w_j-w_{j+2})+
\lambda_j-\lambda_{j+1}=0,\qquad j=1,\ldots,N-1,\la{aaa3}\ee
in system \gl{aaa2} .
In addition, supplementing system  \gl{aaa3}  with the equation
\be w'_N+w^2_N+\lambda_N-V_1=0\la{aaa4}\ee
({\it i.e.}, the Schr\"odinger equation for $W_N$ rewritten for $w_N$) and
summing up the last  $N-n+1$ equations of the new system, we get the relations
\be w'_n+w'_{n+1}+(w_n-w_{n+1})^2+\lambda_n-V_1=0,\qquad n=1,\ldots,N.\ee
\section{Parametric formulas for partial Wronskians}

If $N=3$ system \gl{aaa3}, \gl{aaa4} takes the following form
\be w'_1-w'_3+w_1^2-w_3^2-2w_2(w_1-w_3)+\lambda_1-\lambda_2=0,\la{pr1}\ee
\be w'_2+w_2^2-2w_3w_2+\lambda_2-\lambda_3=0,\la{pr2}\ee
\be w'_3+w_3^2-V_1+\lambda_3=0.\la{pr3}\ee
Let us introduce the function
\be G(x)={1\over2}[w'_1(x)+w_1^2(x)-V_1(x)+\lambda_1+\lambda_2+\lambda_3].
\la{pr4}\ee
Equalities \gl{pr1} and \gl{pr3} imply the identity
\be w_2(w_1-w_3)=G-\lambda_2 .\la{pr5}\ee

In order to derive a formula which expresses $w_3$ in terms of $G$, let us
compare two expressions for $w'_2$: the expression, obtained by
differentiation of the equality
\be w_2={{G-\lambda_2}\over{w_1-w_3}},\la{pr6}\ee
which follows from \gl{pr5}, and the expression deduced from \gl{pr2} after a substitution
of \gl{pr6} into \gl{pr2} instead of $w_2$. By solving the appearing quadratic equation for $w_3$, we come to the equality
\be w_3=w_1+{{G'-\sqrt{(G')^2+4P_3(G)}}\over{2(G-\lambda_3)}},\qquad
P_3(\lambda)\equiv(\lambda-\lambda_1)(\lambda-\lambda_2)(\lambda-\lambda_3)
\la{pr7}\ee
for a certain branch of $\sqrt{(G')^2+4P_3(G)}$. It follows from
\gl{pr5} and \gl{pr7} that
\be w_2={{G'+\sqrt{(G')^2+4P_3(G)}}\over{2(G-\lambda_1)}}.\la{pr9}\ee
Note that formulas \gl{pr6}, \gl{pr7} and \gl{pr9} are valid
if $G(x)$ is different from an identical constant that equal
one of the numbers $\lambda_j$. Below, we show that
the latter condition always takes place  for an intertwining operator
$q_3^-$ that cannot be stripped-off.

Now we derive  formulas which express $w_1$ in terms of $G$.
We substitute \gl{pr7} into \gl{pr1} instead of $w_3$ and obtain
$w_1$ from the resulting expression to deduce that if
\be[G'(x)]^2+4P_3(G(x))\not\equiv0\la{pr10}\ee
then the equality
\be w_1={{G''+2P'_3(G)}\over{2\sqrt{(G')^2+4P_3(G)}}}\la{pr11}\ee
holds; if for some interval
\be[G'(x)]^2+4P_3(G(x))\equiv0,\qquad G'(x)\not\equiv0,\la{pr12}\ee
on some interval, then
\be w_1(x)\equiv0\la{pr13}\ee
on this interval, and the potentials $V_1(x)$ and $V_2(x)$  are identical by \gl{v2v1}.

\section{Smoothness of potentials and coefficients of intertwining
operators}

The following lemma indicates  how smooth are the $V_2(x)$ and coefficients of the 
intertwining operators $q_3^\pm$ for a given smoothness of $V_1(x)$ .

{\bf Lemma 1.} {\it Assume that: 1) $h^\pm=-\partial^2 +V_{1,2}(x)$,
$V_1(x)\in C^n_{\mathbb R}$, $n\geqslant3$, and  $V_2(x)\in C_{\mathbb R}$;

2)
\be q_3^-=\partial^3+\alpha(x)\partial^2+\beta(x)\partial+\gamma(x);\qquad
\alpha(x), \beta(x), \gamma(x)\in C^2_{\mathbb R};\ee

3) $q_3^-$ intertwines $h^+$ and $h^-$, so that
\be q_3^-h^+=h^-q_3^-.\la{ps1}\ee

Then: 

1) $\alpha (x)\in C^{n+1}_{\mathbb R}$, $\beta(x)\in C^n_{\mathbb R}$,
$\gamma(x)\in C^{n-1}_{\mathbb R}$, and $V_2(x)\in C^n_{\mathbb R}$;

2) the operators $q_3^-$ and $q_3^+=(q_3^-)^t$ can be presented in the form
\be
q_3^\mp=\pm\partial^3+g_2(x)\partial^2+[g'_2(x)\mp2g_1(x)]\partial+
[g_0(x)\mp g'_1(x)],\la{p1}\ee

where $$g_2(x)=\alpha(x),\  g_1(x)=[\alpha'(x)-\beta(x)]/2,\ 
g_0(x)=\gamma(x)+[\alpha''(x)-\beta'(x)]/2;$$ 

in addition $g_2(x)\in
C^{n+1}_{\mathbb R}$, $g_1(x)\in C^{n+1}_{\mathbb R}$, $g_0(x)\in C^{n-1}_{\mathbb
R}$;

3) $q_3^+$ intertwines $h^+$ and $h^-$,  so that
\be h^+q_3^+=q_3^+h^-.\la{sp*}\ee}

{\it Proof.} Let us check first that
\be\alpha(x)\in C^{n-1}_{\mathbb R},\qquad \beta(x)\in C^{n-1}_{\mathbb R},\qquad
\gamma(x)\in C^{n-1}_{\mathbb R}.\la{p1'}\ee
Indeed, inclusions \gl{p1'} follow from relations \gl{l11} and \gl{rjrn}
for $j=1$ and $N=3$, from the fact that $\phi_1(x)$, $\phi_2(x)$ and
$\phi_3(x)$ belong to $C^{n+2}_{\mathbb R}$ as formal eigenfunctions and (possibly) associated functions of
$h^+$,  and from the fact that $W_1(x)$,
a Wronskian of basis elements  in ${\rm{ker}}\,q_3^-$, does not have zeroes.

From the intertwining condition \gl{ps1}, we derive the following system of equations:
\be V_2(x)-V_1(x)=2\alpha'(x), \la{p2}\ee
\be\alpha(x)[V_2(x)-V_1(x)]-3V'_1(x)=\alpha''(x)+2\beta'(x),\la{p3}\ee
\be\beta(x)[V_2(x)-V_1(x)]-2\alpha(x)V'_1(x)-3V''_1(x)=
\beta''(x)+2\gamma'(x),\la{p4}\ee
\be\gamma(x)[V_2(x)-V_1(x)]-\beta(x)V'_1(x)-\alpha(x)V''_1(x)-V'''_1(x)=
\gamma''(x).\la{p5}\ee
Relations \gl{p1'}, \gl{p2} and condition 1 imply that $V_2(x)\in
C^{n-2}_{\mathbb R}$. We deduce from \gl{p1'}, \gl{p3}, condition 1, and  the
inclusion $V_2(x)\in C^{n-2}_{\mathbb R}$ that $\alpha''(x)\in
C^{n-2}_{\mathbb R}$, {\it i.e.}, $\alpha(x)\in C^n_{\mathbb R}$. From \gl{p2},
condition 1 and the fact that $\alpha(x)\in C^n$ it follows
that $V_2(x)\in C^{n-1}_{\mathbb R}$. It follows from \gl{p1'},
\gl{p4}, condition 1 and the inclusions  $V_2(x)\in C^{n-1}_{\mathbb R}$
and  $\alpha(x)\in C^n_{\mathbb R}$ that $\beta(x)\in C^n_{\mathbb R}$. From
\gl{p3}, condition 1, and the inclusions $V_2(x)\in C^{n-1}_{\mathbb
R}$, and $\alpha(x)$ and $\beta(x)\in C^n_{\mathbb R}$ it follows that
$\alpha(x)\in C^{n+1}_{\mathbb R}$. Finally, it follows  from \gl{p2}, 
condition 1 and the inclusion $\alpha(x) \in C^{n+1}_{\mathbb R}$ that
$V_2(x)\in C^n_{\mathbb R}$. Thus, the first statement is proved.

The validity of equality \gl{p1} is easily verified with the help of straightforward
calculations. The fact that  $g_2(x)$ and $g_0(x)$ are in $C^{n+1}_{\mathbb R}$ and
$C^{n-1}_{\mathbb R}$, respectively, and equality \gl{sp*} are obvious. Finally,
to show that $g_1(x)$ belongs to $C^{n+1}_{\mathbb R}$ we refer to the equalities
\be 3V_1(x)+\alpha'(x)+2\beta(x)\in C^{n+1}_{\mathbb R},\la{p6}\ee
\be 3V'_1(x)+\beta'(x)+2\gamma(x)\in C^n_{\mathbb R},\la{p7}\ee
and
\be V'_1(x)+\gamma(x) \in C^n_{\mathbb R},\la{p9}\ee
which, in turn, follow  from equalities \gl{p3}--\gl{p5} since
$V_{1,2}(x)\in C^n_{\mathbb R}$, $\alpha(x)\in C^{n+1}_{\mathbb R}$, $\beta(x)
\in C^n_{\mathbb R}$ and $\gamma(x)\in C^{n-1}_{\mathbb R}$. Lemma 1 is proved.

{\bf Corollary 3.} By calculating  the coefficient at $\partial^2$ in $q_3^-$ with the help of \gl{l11}, \gl{l12} and
\gl{aaasuper}, we deduce that
\be\alpha(x)\equiv g_2(x)\equiv-{{W'_1(x)}\over{W_1(x)}}\equiv-w_1(x).\la{pr14}\ee
Hence, under the conditions of Lemma 1
\be w_1(x)\in C^{n+1}_{\mathbb R},\quad\mbox{and}\quad W_1(x)=Ce^{\int w_1(x)\,dx}
\in C^{n+2}_{\mathbb R}.\la{pr15}\ee

\section{Parametric formulas for coefficients of intertwining operators}

It was shown in \cite{iofnis} that the potentials $V_1(x)$, $V_2(x)$ and
coefficients of the intertwining operator $q_3^+$ can be parameterized by a single
function which was denoted $W(x)$ in \cite{iofnis}. It is not difficult to check that
this function is connected with $G(x)$ by the relation
\be W=G-{1\over3}(\lambda_1+\lambda_2+\lambda_3).\la{pr17}\ee
For convenience, we give parametric formulas obtained in
\cite{iofnis} in the notation of the present work,
\be V_{1,2}=g_2^2\mp g'_2-2G+\lambda_1+\lambda_2+\lambda_3,\la{pr18}\ee
\be g_1={1\over2}[g_2^2-3G+\lambda_1+\lambda_2+\lambda_3],\la{pr19}\ee
\be g_0=g''_2-g_2[g_2^2-3G+\lambda_1+\lambda_2+\lambda_3]+
{1\over2}\sqrt{(G')^2+4P_3(G)}.\la{pr20}\ee
Let us emphasize that, in contrast to
\gl{pr11}, parameterizations \gl{pr18}--\gl{pr20}  are valid for any case. Parameterization \gl{pr11} which supplements
\gl{pr18}--\gl{pr20}, (see, in addition, \gl{pr14}) is derived in \cite{iofnis}
as well, but only under the following conditions
\be G(x)\ne{\rm{Const}},\qquad[G'(x)]^2+4P_3(G(x))\not\equiv0,\la{pr21}\ee
whereas in the present work we derive this parameterization under a weaker condition \gl{pr10}.
The case of conditions \gl{pr12}, \gl{pr13} was not considered in \cite{iofnis}.

\section{Relations between parameterization function and partial Wronskians}

In the following lemma, we indicate basic relations between the parameterization function $G(x)$ and Wronskians of
a part of canonical basis elements in the kernel of $q_3^-$.

{\bf Lemma 2.} {\it Assume that: 1) the conditions of Lemma 1 are
fulfilled; 2) $W_{jk}=\phi'_j
\phi_k-\phi_j\phi'_k$; 3) $\sqrt{(G')^2+4P_3(G)}$ is the same branch of the
root as above. Then:

1) if a Jordan form of the matrix $\bf S$ for the operator $q_3^-$ contains
three Jordan cells of  first order,
\be h^+\phi_1=\lambda_1\phi_1,\qquad h^+\phi_2=\lambda_2\phi_2,\qquad
h^+\phi_3=\lambda_3\phi_3,\la{pr21'}\ee
then the following identities hold:
\be W'_{kl}=(\lambda_l-\lambda_k)\phi_k\phi_l,\qquad\Big({{\phi_j}\over{W_1}}
\Big)'=\varepsilon_{jkl}(\lambda_k-\lambda_l){{W_{jk}W_{jl}}\over{W_1^2}},
\la{pr22}\ee
\be G-\lambda_j=\varepsilon_{jkl}(\lambda_j-\lambda_k)(\lambda_j-\lambda_l)
{{\phi_jW_{kl}}\over{W_1}},\la{pr23}\ee
\be G'+\sqrt{(G')^2+4P_3(G)}=2(\lambda_1-\lambda_2)(\lambda_2-\lambda_3)
(\lambda_3-\lambda_1){{\phi_1\phi_2\phi_3}\over W_1},\la{pr24}\ee
\be G'-\sqrt{(G')^2+4P_3(G)}=2(\lambda_1-\lambda_2)(\lambda_2-\lambda_3)
(\lambda_3-\lambda_1){{W_{12}W_{23}W_{31}}\over W_1^2},\la{pr25}\ee
where $j,k,l$ is an arbitrary permutation of $1,2,3$ and summation for a
repeated index is not performed; in addition, the branch of
$\sqrt{(G')^2+4P_3(G)}$ is independent of the choice of numbering of
canonical basis elements;

2) if a Jordan form of the matrix $\bf S$ for the operator $q_3^-$ contains a
Jordan cell of first order and a Jordan cell of second order:
\be h^+\phi_1=\lambda_1\phi_1,\qquad h^+\phi_2=\lambda_2\phi_2+\phi_3,\qquad
h^+\phi_3=\lambda_3\phi_3,\qquad\lambda_3=\lambda_2,\la{pr26}\ee
then the following identities hold:
\be W'_{13}=(\lambda_2-\lambda_1)\phi_1\phi_3,\quad W'_{23}=-\phi_3^2,
\qquad\Big({{\phi_1}\over{W_1}}\Big)'={{W_{13}^2}\over{W_1^2}},
\quad\Big({{\phi_3}\over{W_1}}\Big)'=(\lambda_1-\lambda_2){{W_{13}W_{23}}
\over{W_1^2}},\la{pr28}\ee
\be G-\lambda_1=(\lambda_1-\lambda_2)^2{{\phi_1W_{23}}\over{W_1}},
\qquad G-\lambda_2=(\lambda_1-\lambda_2){{\phi_3W_{13}}\over{W_1}},\la{pr29}\ee
\be G'+\sqrt{(G')^2+4P_3(G)}\!=\!-2(\lambda_1-\lambda_2)^2
{{\phi_1\phi_3^2}\over W_1},
\,G'-\sqrt{(G')^2+4P_3(G)}\!=\!2(\lambda_1-\lambda_2)^2
{{W_{13}^2W_{23}}\over W_1^2},\la{pr30}\ee
and the branch of $\sqrt{(G')^2+4P_3(G)}$ is independent of the choice of numbering
of canonical basis elements;

3) if the Jordan cell of the matrix $\bf S$ for the operator $q_3^-$
consists of a single Jordan cell of  third order:
\be h^+\phi_1=\lambda_1\phi_1+\phi_2,\qquad h^+\phi_2=\lambda_2\phi_2+\phi_3,
\qquad h^+\phi_3=\lambda_3\phi_3,\qquad\lambda_3=\lambda_2=\lambda_1,\la{pr31}\ee
then the following identities hold:
\be W'_2=-\phi_3^2,\qquad\Big({{\phi_3}\over{W_1}}\Big)'={{W_2^2}\over{W_1^2}},
\la{pr32}\ee
\be G-\lambda_1={{\phi_3W_2}\over{W_1}},\la{pr33}\ee
\be G'+\sqrt{(G')^2+4P_3(G)}=-2{{\phi_3^3}\over W_1},
\qquad G'-\sqrt{(G')^2+4P_3(G)}=2{{W_2^3}\over W_1^2}.\la{pr34}\ee}

{\bf Proof.} Identities \gl{pr22}, \gl{pr28} and \gl{pr32} are easily
checked with the help of straightforward calculations in which we 
use relations \gl{pr21'}, \gl{pr26} and \gl{pr31}.

Identities \gl{pr23}, \gl{pr29} and \gl{pr33} follow from identity
\gl{pr5} - for convenience, we write the latter identity in the case considered in the form
\be G-\lambda_2=-{{W'_{23}}\over W_{23}}{{(\phi_3/W_1)'}\over{\phi_3/W_1}};
\la{pr35}\ee
as well, those identities follow from identities \gl{pr22}, \gl{pr28}, \gl{pr32} and from the fact that we can renumber elements of
canonical basis so that any given eigenvalue of the
matrix $\bf S$ for the operator $q_3^-$ gets index 2 (see \gl{pr35}).

In the case
$G\not\equiv\lambda_j$, $j=1,2,3$  identities \gl{pr24}, \gl{pr25}, \gl{pr30}, and \gl{pr34}  follow from identities \gl{pr23}, \gl{pr29}
and \gl{pr33} and relations \gl{pr7}, \gl{pr9}. Before we prove identities
\gl{pr24}, \gl{pr25}, \gl{pr30} and \gl{pr34} in the case $G\equiv\lambda_j$,
$j=1,2,3$ let us show that this case is equivalent to the existence of two
Jordan cells with the same eigenvalue $\lambda_j$ in a Jordan form of the matrix
$\bf S$ for the operator $q_3^-$. In the latter case, the validity of the desired identities 
is obvious.

 Identities
\gl{pr22}, \gl{pr28}, \gl{pr32} and the fact, that any formal eigenfunction
of the Hamiltonian (different from identical zero) can have zeroes of  first
order only imply that the right-hand sides of expressions \gl{pr23}, \gl{pr29} and \gl{pr33} either are identical zeroes (if there are two Jordan cells with the
same eigenvalue in a Jordan form of the matrix $\bf S$ for the operator $q_3^-$) or can have zeroes of the order not exceeding four. Thus,
the identity $G\equiv\lambda_j$ holds on some interval if and only if this identity holds on whole axis which is equivalent to the existence of two Jordan cells with the
same eigenvalue $\lambda_j$ in a Jordan form of the matrix $\bf S$ for the operator
$q_3^-$. Hence, identities \gl{pr24}, \gl{pr25}, \gl{pr30} and \gl{pr34}
are proved.

To show that the branch of  $\sqrt{(G')^2+4P_3(G)}$ is independent of the numbering of
canonical basis elements, one can renumber
these elements, derive  formulas similar to \gl{pr24}, \gl{pr25},
\gl{pr30} and \gl{pr34} for new numbering and compare the results. Lemma 2
is proved.

{\bf Corollary 4.} In the proof of Lemma 2, it was shown that for any $j$
the function $G(x)-\lambda_j$ either is identical zero on the whole axis or can
have zeroes only of order not exceeding four. In addition, the relation
$G(x)\equiv\lambda_j$ is equivalent to the existence of two Jordan cells with the
same eigenvalue $\lambda_j$ in a Jordan form of the matrix $\bf S$ for the operator
$q_3^-$, which in view of Theorem 2, is equivalent  to the possibility to strip-off the
operators $q_3^\mp$.

{\bf Corollary 5.} Under the conditions of Lemma 1, the functions
$\phi_1(x)$, $\phi_2(x)$ and $\phi_3(x)$ belong to $C^{n+2}_{\mathbb R}$; hence it follows
from identities \gl{pr22}, \gl{pr28} and \gl{pr32}, that
the Wronskians $W_{kl}(z)$ (case 1), $W_{13}(x)$ and $W_{23}(x)$ (case 2), and
$W_2(x)$ (case 3) belong to $C^{n+3}_{\mathbb R}$. We apply these inclusions, take
differences of identities \gl{pr24} and \gl{pr25}; \gl{pr30};
\gl{pr34}, refer to inclusions \gl{pr15},  and take into account that $W_1(x)$ has no zeroes 
(as the Wronskian of a basis in ${\rm{ker}}\,q_3^-$) to show that
\be\sqrt{(G')^2+4P_3(G)}\in C^{n+2}_{\mathbb R}.\la{pr36}\ee
In its turn, inclusion \gl{pr36} and formulas \gl{pr24}, \gl{pr30} and \gl{pr34}   provide that
\be G(x)\in C^{n+3}_{\mathbb R}.\la{pr36'}\ee

{\bf Corollary 6.} If coefficients of $q_3^-$ are real, then coefficients of
the polynomial $P_3(h^\pm)=q_3^\pm q_3^\mp$ are real as well. Hence, either all of the
numbers $\lambda_j$ are real or  one of these numbers is real and two are
mutually complex conjugate. Without loss of generality, we assume
that elements of the canonical basis in ${\rm{ker}}\,q_3^-$ that correspond to real
$\lambda_j$, are chosen  real, and elements that correspond to complex conjugate $\lambda_j$ are
complex conjugate. Then, if all of $\lambda_j$ are real, the root
$\sqrt{(G')^2+4P_3(G)}$  is real 
for any $x\in\mathbb R$ in view of \gl{pr24}, \gl{pr30} and \gl{pr34}. If there is pair of complex conjugate values
$\lambda_j$ (obviously,  this is possible  only if all of $\lambda_j$
are different), then $W_1(x)$ is purely imaginary (since complex conjugation
of $W_1(x)$ corresponds to a permutation of two lines in the definition of $W_1(x)$)
and $\sqrt{(G')^2+4P_3(G)}$  is also real for any
$x\in\mathbb R$ by virtue of \gl{pr24}. Thus,
\be[G'(x)]^2+4P_3(G(x))\geqslant0,\qquad x\in\mathbb R.\la{pr37}\ee

\section{Lower bound of the parameterization function}

A lower bound for the parameterization function $G(x)$ is given by the following lemma.

{\bf Lemma 3.} {\it If, under the conditions of Lemma 1, the intertwining operator
$q_3^-$ cannot be stripped-off, coefficients of $q_3^-$ are real and
$\lambda_3$ is the minimal real eigenvalue of the matrix $\bf S$ for the operator $q_3^-$,
then the inequality 
\be G(x)>\lambda_3,\qquad x\in\mathbb R,\la{gb}\ee
holds.}
{\bf Proof.} First we  show that the inequality
\be G(x)\geqslant\lambda_3,\qquad x\in\mathbb R\la{gbr}\ee
takes place. Assume that for some point $x_0\in\mathbb R$,  inequality
\gl{gbr} is violated. Then  $P_3(G(x_0))<0$ and, consequently,
by  \gl{pr37} ,the derivative $G'(x_0)\ne0$. Moreover, $G'(x)$
does not vanish on the entire interval which contains $x_0$ and on which the inequality
$G(x)<\lambda_3$ holds. Hence, $G(x)$  either strictly increases or
strictly decreases on this interval. Obviously the interval is not bounded  from the
left (right), if $G(x)$  increases (decreases) on it. Let us show that the
assumption about the violation of \gl{gbr} leads to a contradiction. For definiteness, we 
consider  the case where $G(x)$ increases on the above-mentioned
interval. The case of decreasing $G(x)$ is treated similarly. By
inequality \gl{pr37},  the inequality
$G'(x)/\sqrt{-P_3(G(x))}\geqslant2$ is valid for any point of the considered interval. Integrating the latter inequality
from $x$ to $x_0$, we deduce that
\be \int\limits_{G(x)}^{G(x_0)}{{dG}\over\sqrt{-P_3(G)}}\geqslant2(x_0-x),
\qquad
x<x_0. \la{100}\ee
The left-hand side of inequality \gl{100} is bounded for $x\to-\infty$ while its
right-hand side tends to $+\infty$. This contradiction proofs inequality
\gl{gbr}.

To prove that $G(x)-\lambda_3$  has no zeroes, we use identity
\gl{pr11} which expresses $g_2(x)$ in terms of $G(x)$ (see also \gl{pr14}).
Let us assume that there is a point $x_0\in \mathbb R$ such that $G(x_0)=\lambda_3$.
Since $q_3^-$ cannot be stripped-off, Corollary 4
and inequality \gl{gbr} at the point $x_0$ imply that the function $G(x)-\lambda_3$ has a
zero of  even order $2n$, $G'(x)$ has a zero of  order $2n-1$, and $G''(x)$
has a zero of  order $2n-2$, where $n$ is either 1 or 2; in addition, it is obvious that
\be G''(x_0)\geqslant0.\la{gvp}\ee

First we consider the case where $\lambda_3$ is a zero of
$P_3(\lambda)$ of order one. In this case, inequality \gl{gbr} and the fact that the order of the root of 
$G(x)-\lambda_3$  is even imply condition \gl{pr10}, which allows us to use formula \gl{pr11}. Finally, since 
\be P'_3(G(x_0))=(\lambda_3-\lambda_1)(\lambda_3-\lambda_2)>0,\ee
and inequality \gl{gvp} holds, the right-hand side of \gl{pr11} at the point $x_0$
is infinite, which contradicts to \gl{pr15}. Hence, $G(x)$ cannot 
equal  $\lambda_3$.

Now we assume that $\lambda_3$ is a zero of $P_3(\lambda)$ of order two or three. In this case, 
the numerator of \gl{pr11} has, obviously, a zero of  order
$2n-2$ at the point $x_0$ and the denominator has a zero of  order $2n-1$. Hence, $g_2(x)$ has
a pole at the point $x_0$, which is impossible. Thus, $G(x)$ cannot equal
$\lambda_3$, and Lemma 3 is proved.

\section{Theorem on reducibility of intertwining operators of the
third order}

The assertion that any intertwining operator of the third order with real
coefficients is really reducible is described by the following theorem.\\
{\bf Theorem 5.} {\it Assume that the conditions of Lemma 1 are satisfied and that the intertwining operator 
$q_3^-$ cannot be stripped-off has real coefficients. Let $\lambda_3$
be the minimal real eigenvalue of the matrix $\bf S$ for the operator $q_3^-$. Then
there exist intertwining operators  $p_{1}^\pm$ and $k_{1}^\pm$ of the first order
and  $p_{2}^\pm$ and $k_{2}^\pm$ of the second order such that:\\

1) coefficients of $p_{1}^\pm$ and $k_{1}^\pm$ are real, and, in addition,
coefficients of these operators at $\partial^0$ belong to $C^{n+1}_{\mathbb R}$;

2) coefficients of $p_{2}^\pm$ and $k_{2}^\pm$ are real, and, in addition,
coefficients of these operators at $\partial$ and $\partial^0$ belong
to $C^{n+2}_{\mathbb R}$ and $C^n_{\mathbb R}$, respectively;

3)
\be p_1^+=(p_1^-)^t,\qquad k_1^+=(k_1^-)^t,\qquad
p_2^+=(p_2^-)^t,\qquad k_2^+=(k_2^-)^t;\ee

4) the matrices $\bf S$ for the operators $p_{1}^\pm$ and $k_{1}^\pm$ consist of
$\lambda_3$;

5)
\be q_3^-=k_{2}^-p_{1}^-=k_{1}^-p_{2}^-,\qquad
q_3^+=p_{1}^+k_{2}^+=p_{2}^+k_{1}^+,\la{pfact}\ee
\be p_{1}^-h^+=h_1p_{1}^-,\qquad k_{2}^-h_1=h^-k_{2}^-,\qquad
p_{2}^-h^+=h_2p_{2}^-,\qquad k_{1}^-h_2=h^-k_{1}^-,\la{pr50}\ee
\be h^+p_{1}^+=p_{1}^+h_1,\qquad h_1k_{2}^+=k_{2}^+h^-,\qquad
h^+p_{2}^+=p_{2}^+h_2,\qquad h_2k_{1}^+=k_{1}^+h^-,\la{pspl}\ee
where $h_1$ and $h_2$ are intermediate Hamiltonians with real potentials
from $C^n_{\mathbb R}$.}

{\bf Proof.} We consider the case with $p_{1}^-$ and $k_{2}^-$ only
since the statements of Theorem 5  for the case of $p_{1}^+$
and $k_{2}^+$ are easily verifiable with the help of transposition, and the statement for
the cases of $k_{1}^\pm$ and $p_{2}^\pm$ follows from the symmetry between
$h^+$ and $h^-$.

Let us define $p_{1}^-$ and $k_{2}^-$ by the equalities
\be p_{1}^-=r_3^-,\qquad k_{2}^-=r_1^-r_2^-.\ee
Then existence of an intermediate Hamiltonian $h_1$ and intertwining \gl{pr50}
follows from relations \gl{aaaprom} and \gl{aaaspl}. The fact that the  potential of the Hamiltonian $h_1$ given
 by the formula
\be V_1(x)-2[\ln\phi_3(x)]''\equiv V_1(x)-2w'_3(x)\ee
(see \gl{v2v1}) is real and belongs to the
space $C^n_{\mathbb R}$
follows from \gl{pr7}, \gl{pr14}, \gl{pr37}, \gl{gb}, from the fact that  $g_2(x)$ is real, 
and from the inclusions \gl{pr36}, \gl{pr36'}, and $g_2(x)\in C^{n+1}_{\mathbb R}$
(see Lemma 1). In addition, the
function
\be\chi_3(x)\equiv -{\phi'_3\over\phi_3}\equiv-w_3,\ee
which is the coefficient at $\partial^0$ of the operator $p_{1}^-$, is obviously  real and belongs to $C^{n+1}_{\mathbb R}$ . To prove that coefficients of $k_{2}^-$ are real and belong to the 
spaces of smooth functions of Theorem 5 we first apply relations \gl{l12}, \gl{aaasuper}, \gl{pr2},
\gl{pr3} and \gl{pr5}  to transform $k_{2}^-$ to the form
\be k_{2}^-=\partial^2+(w_3-w_1)\partial+(G+V_1-w_3^2-w_1w_3-2\lambda_3),\ee
and then take into account the following statements:
 $V_1(x)$, $w_1(x)\equiv-g_2(x)$, $\lambda_3$ (see the Theorem 5
conditions), $G(x)$ and $w_3(x)$ are real, identity \gl{pr7} and inclusions
 \gl{pr15}, \gl{pr36}, and \gl{pr36'} hold, $w_3(x)\in C^{n+1}_{\mathbb R}$,
and $V_1(x)\in C^n_{\mathbb R}$. Finally, the fact that the matrix $\bf S$ for the
operator $p_{1}^-$ consists of $\lambda_3$ follows from \gl{aaaprom}.
Theorem 5 is proved.\\

The work was supported by the RFBR Grant 06-01-00186-a. The first author was supported
 by the Programs ``Development of scientific potential of higher
school'', grant RPN 2.1.1.1112 and ``Leading scientific schools of Russia'',
grant LSS 2.1.1.1112.


\begin{thebibliography}{99}
\bibitem{coop}
F. Cooper and B. Freedman, {\it Ann. Phys. (NY)}, {\bf 146}, 262 (1983).
\bibitem{abi1} A. A. Andrianov, N. V. Borisov, and M. V. Ioffe, {\it
JETP Lett.},{\bf 39}, 93 (1984);\,
{\it Phys. Lett. A}, {\bf 105}, 19 (1984);
{\it Theor. Math. Phys.}, {\bf 61},183 (1984).
\bibitem{miel} B.  Mielnik, {\it
 J.Math.Phys.}, {\bf 25}, 3387 (1984).
\bibitem{mnieto}
M. M. Nieto, {\it  Phys. Lett. B}, {\bf 145}, 208 (1984).
\bibitem{fern} D. J. C.
Fern\'andez, {\it Lett. Math. Phys.}, {\bf 8}, 337 (1984) [physics/0006119].
\bibitem{genkri} L. E. Gendenshtein and I. V. Krive, {\it Sov. Phys. Usp.},  {\bf 28}, 645 (1985).
\bibitem{lahiri}
A. Lahiri, P. K. Roy, and B. Bagchi, {\it Int. J. Mod. Phys. A},
{\bf 5}, 1383 (1990).
\bibitem{cooper95}
F. Cooper, A. Khare, and U. Sukhatme, {\it Phys. Rept.},  {\bf 251}, 267 (1995).
\bibitem{junker}
G. Junker,  {\it Supersymmetric Methods in Quantum and
Statistical Physics}, Springer, Berlin (1996).
\bibitem{bagsam1}
V. G. Bagrov and B. F. Samsonov, {\it Phys. Part. Nucl.},  {\bf 28}, 374 (1997).
\bibitem{spirid1}
V. P. Spiridonov, {\it The factorization method, self-similar potentials and quantum algebras}.  hep-th/0302046.
\bibitem{ancan} A. A. Andrianov and F. Cannata, {\it J. Phys. A: Math. Gen.}, {\bf 37}, 10297 (2004).
\bibitem{darb} G. Darboux, {\it  C. R. Acad. Sci. (Paris)}, {\bf 94}, 1456 (1882) [physics/9908003].
\bibitem{crum} M. M. Crum, {\it
Quart. J. Math. ( Oxford)}, {\bf 6} 121 (1955) [physics/9908019].
\bibitem{krein}
M. G. Krein, {\it Dokl. Akad. Nauk SSSR}, {\bf 113}, 970 (1957).
\bibitem{fadd}
L. D. Faddeev, {\it  Usp. Mat. Nauk}, {\bf 14}, 57 (1959)
[{\it J. Math. Phys.},  {\bf 4}, 72 (1963) ].
\bibitem{matv}
V. B. Matveev and M. Salle,
{\it Darboux transformations and solitons}.   Springer, Berlin (1991).
\bibitem{schr}
E. Schr\"odinger, {\it
Proc. Roy. Irish Acad. A},  {\bf 47}, 53 (1941) [physics/9910003].
\bibitem{inf}
L. Infeld  and T. E. Hull, {\it Rev. Mod. Phys.},  {\bf 23}, 21 (1951).
\bibitem{abei} A. A. Andrianov, N. V. Borisov, M. V. Ioffe, and M. I. Eides, {\it
Theor. Math. Phys.}, {\bf 61}, 17 (1985);
{\it Phys. Lett. A}, {\bf 109}, 143 (1985).
\bibitem{suku} C. V. Sukumar, {\it J. Phys. A: Math. Gen.}, {\bf 18}, L57; 2917 (1985).
\bibitem{ais} A. A. Andrianov, M. V. Ioffe, and V. P. Spiridonov, {\it
Phys. Lett. A}, {\bf 174}, 273 (1993).
\bibitem{acdi95}
A. A. Andrianov, F. Cannata, J.-P. Dedonder, and M. V. Ioffe, {\it
 Int. J. Mod. Phys. A}, {\bf 10}, 2683 (1995).
\bibitem{bags95} V. G. Bagrov and B. F. Samsonov, {\it
Theor. Math. Phys.}, {\bf 104}, 1051 (1995).
\bibitem{ast0}
A. Aoyama, M. Sato, and T. Tanaka, {\it Nucl. Phys. B },{\bf 619}, 105 (2001).
\bibitem{ansok}
A. A. Andrianov and A. V. Sokolov, {\it
Nucl. Phys. B}, {\bf 660}, 25 (2003).
\bibitem{sam1} B.F. Samsonov,{\it Phys. Lett. A},{\bf 263}, 274 (1999).
\bibitem{fer1} D.J. Fern\'andez C., R. Mu\~noz, and A. Ramos,{\it Phys. Lett. A},{\bf 308},11  (2003).
\bibitem{fer2} D.J. Fern\'andez C. and E. Salinas-Hern\'andez, {\it J. Phys. A: Math. Gen.},{\bf 36}, 2537 (2003).
\bibitem{sam2} B.F. Samsonov and F. Stancu, {\it Phys. Rev. C},{\bf 67}, 054005 (2003).
\bibitem{sam3} B.F. Samsonov, {\it Phys. Lett. A},{\bf 358}, 105 (2006).
\bibitem{naim} M. A. Naimark, {\it Linear differential operators}. 
Frederick Ungar Publishing Co., New York (1967).
\bibitem{acin1} A. A. Andrianov, F. Cannata, M. V. Ioffe, and D. N. Nishnianidze. {\it
 Phys. Lett. A}, {\bf 266}, 341 (2000).
\bibitem{iofnis} M. V. Ioffe and
  D. N. Nishnianidze, {\it  Phys. Lett. A}, {\bf 327}, 425 (2004).
\end{thebibliography}
\end{document}